\documentclass[conference]{IEEEtran}
\IEEEoverridecommandlockouts
\usepackage{cite}
\usepackage{amsmath,amssymb,amsfonts}
\usepackage{algorithmic}
\usepackage{graphicx}
\usepackage{textcomp}
\usepackage{xcolor}

\usepackage{subcaption}
\usepackage{multirow}
\usepackage[inline]{enumitem}
\usepackage{gensymb}

\usepackage{authblk}
\usepackage[a4paper, total={184mm,239mm}]{geometry}
\def\BibTeX{{\rm B\kern-.05em{\sc i\kern-.025em b}\kern-.08em
T\kern-.1667em\lower.7ex\hbox{E}\kern-.125emX}}

\begin{document}

\title{NeuroAI Temporal Neural Networks (NeuTNNs): Microarchitecture and Design Framework for Specialized Neuromorphic Processing Units}

\author{
Shanmuga Venkatachalam, Prabhu Vellaisamy, Harideep Nair, Wei-Che Huang, Youngseok Na, Yuyang Kang, \\Quinn Jacobson, and John Paul Shen}
\affil{Carnegie Mellon University}

\maketitle

\begin{abstract}
Leading experts from both communities have suggested the need to
(re)connect research in neuroscience and artificial intelligence (AI) to accelerate the development of next-generation AI innovations. 
They term this convergence as \textit{NeuroAI}. 
Previous research has established temporal neural networks (TNNs) as a promising neuromorphic approach toward biological intelligence and efficiency. We fully embrace NeuroAI and propose a new category of TNNs we call NeuroAI TNNs (\textit{NeuTNNs}) with greater capability and hardware efficiency by adopting neuroscience findings, including a neuron model with active dendrites and a hierarchy of distal and proximal segments. This work introduces a PyTorch-to-layout tool suite (\textit{NeuTNNGen}) to design application-specific NeuTNNs. Compared to previous TNN designs, NeuTNNs achieve superior performance and efficiency. We demonstrate NeuTNNGen's capabilities using three example applications: 1) UCR time series benchmarks, 2) MNIST design exploration, and 3) Place Cells design for neocortical reference frames. 
We also explore using synaptic pruning to further reduce synapse counts and hardware costs by 30-50\% while maintaining model precision across diverse sensory modalities.
NeuTNNGen can facilitate the design of application-specific energy-efficient NeuTNNs for the next generation of NeuroAI computing systems.

\end{abstract}


\section{Background \& Introduction}
\label{sec:intro}

\subsection{Neuromorphic Temporal Neural Networks}

Spiking Neural Networks (SNNs) \cite{maass1997networks} are artificial neural networks that mimic biological neural processes to perform computation. Temporal Neural Networks (TNNs) as proposed by J.E. Smith \cite{smith2017space, smith2018space,smith2020temporal} are a special class of SNNs that encode and process data using spike timings (temporal encoding).
TNNs employ local learning adapted from biology called spike-timing-dependent plasticity (STDP) \cite{bi1998synaptic,nessler2009stdp,kheradpisheh2018stdp,khoee2024meta,rahman2024modulated}. Due to its adherence to biological plausibility and local nature, STDP has several desirable attributes such as high scalability, rapid few-shot learning, and online continual learning, which are difficult to achieve with backpropagation-based learning \cite{rumelhart1985learning} in deep neural networks (DNNs) \cite{lecun2015deep}. TNNs aggressively embrace such brain-like attributes to achieve brain-like energy efficiency.

A microarchitecture model has been proposed for direct hardware implementation of TNNs \cite{nair2020direct}, and further works have demonstrated the feasibility of efficient implementation of TNNs \cite{catwalk25, nair2024nertcam, nair2024tnn}. 
TNN7 \cite{nair2022tnn7}, a custom macro suite of highly optimized TNN building blocks augmenting the ASAP7 process design kit (PDK) \cite{clark2016asap7}, has been proposed to further enhance implementation efficiency and design productivity. 
As a first step \cite{vellaisamy2022towards} toward automating and streamlining the design flow, TNNGen \cite{vellaisamy2024tnngen} demonstrated the PyTorch to layout implementation of single-layer feedforward TNNs for time-series applications. These recent efforts have shown the efficacy of TNNs \cite{shen2023cortical}, demonstrating state-of-the-art clustering and classification performance within tens of $\mu$W power across various time series signal modalities \cite{chaudhari2021unsupervised}, and tens of $m$W power with MNIST digit recognition benchmark \cite{nair2022tnn7,smith2020temporal,smith2023neuromorphic} with the capability for \textit{online continual learning}.

\subsection{Recent Insights from Neuroscience Research}

Significant progress has been made in theoretical and experimental neuroscience to better understand the neocortex. 
Hawkins \cite{hawkins2021thousand} postulates that \textit{Cortical Columns (CCs)} serve as the foundational compute units within the mammalian neocortex (brain's house of intelligence).
CCs leverage structured \textit{Reference Frame (RF)} to model sensory information from diverse sensory modalities (e.g., visual, audio, tactile, etc.) via sensorimotor learning \cite{hawkins2017theory}.
Reference frames (RFs) are directly inspired from the biological framework that humans use to navigate the real world, implemented primarily by entorhinal grid cells \cite{hafting2005microstructure} and hippocampal place cells \cite{o1978hippocampal}. 
%

\begin{figure}[!h]
\centering
\includegraphics[width=1.04\columnwidth]{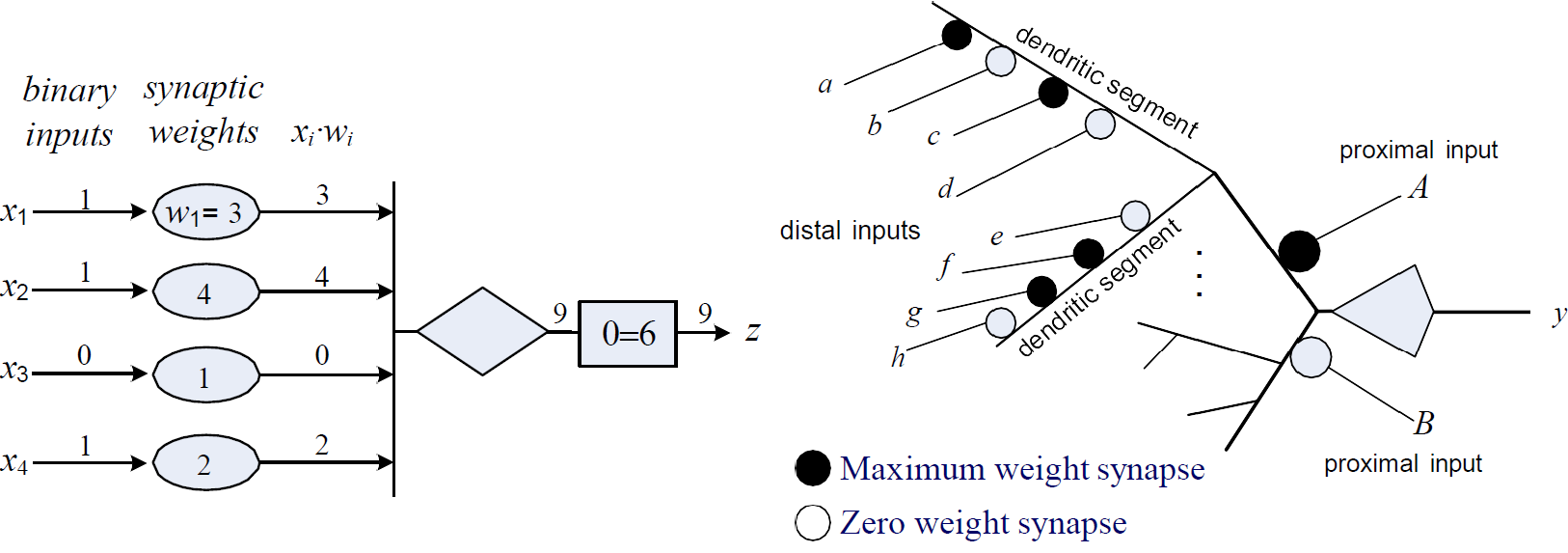} 
\caption{Left is the classic point neuron model, and right is the new neuron model with multiple active dendrites. Each active dendrite can contain multiple segments, some distal and some proximal. Each segment is comparable to a point neuron and able to recognize a specific pattern.
Figure taken from \cite{smith2022macrocolumn}.}
\label{fig:ADneuron}
\end{figure}

CCs and RFs are composed of neurons with \textit{Active Dendrites (AD)} ingesting \textit{distal} and \textit{proximal} inputs. Such an implementation of RF has been demonstrated in \cite{smith2022macrocolumn}.
Fig. \ref{fig:ADneuron} illustrates the neuron model with active dendrites. Proximal inputs have a larger effect on the body potential of the neuron and are necessary for the neuron to fire. Distal inputs provide contextual information via preliminary clustering, enabling the neuron to fire earlier. In effect, AD neurons perform multiple stages of preprocessing and clustering prior to input accumulation \cite{ahmad2016neurons,hole2021thousand}. These additional abstractions make AD neurons much more computationally powerful than point neurons used in prior SNNs and DNNs with a single point of input accumulation into body potential \cite{grewal2021going, iyer2022avoiding}.

\subsection{A New Neuromorphic Processor Design Approach}

\begin{figure}[!h]
\centering
\includegraphics[width=0.8\columnwidth]{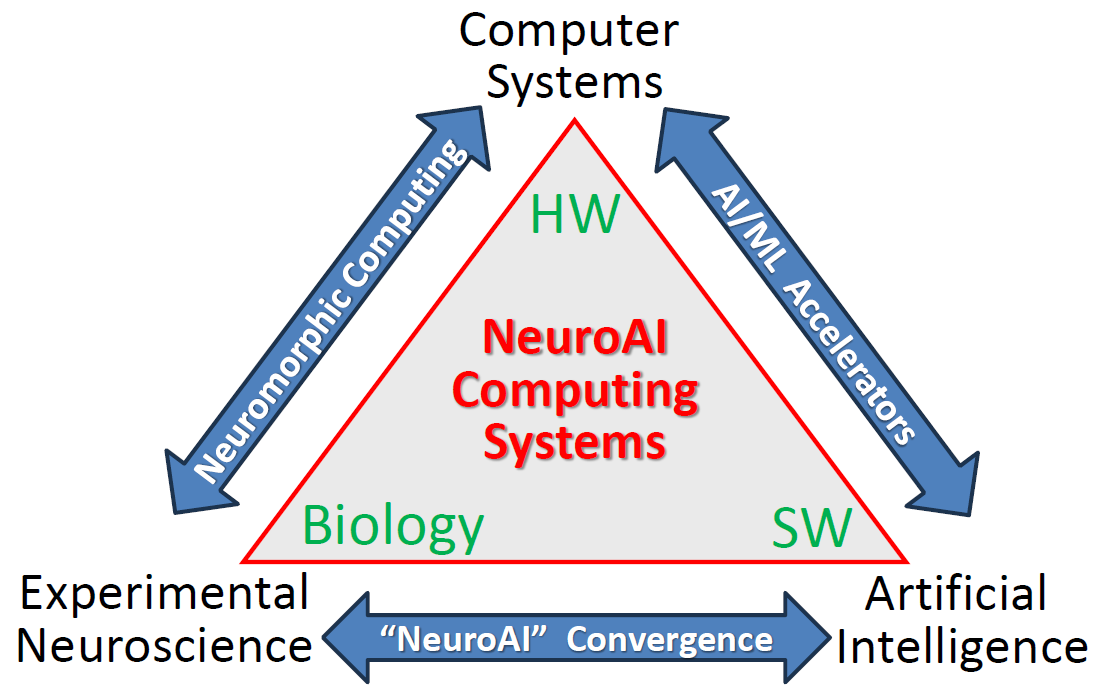} 
\caption{NeuroAI Computing Systems embrace the convergence of Neuroscience, AI, and Computer Systems. We fully support and leverage this convergence in our NeuroAI TNN microarchitecture (NeuTNN) and design framework (NeuTNNGen).}

\label{fig:tripleconv}
\end{figure}

Leading experts have suggested the need to merge neuroscience and artificial intelligence (AI) research \cite{zador2023catalyzing} to accelerate the development of next-generation AI with more human-like capabilities. They term this two-way convergence: \textit{NeuroAI}. We fully embrace NeuroAI and extend it to also include Computer Systems, to build next-generation AI systems with more brain-like capabilities and efficiency (see Fig. \ref{fig:tripleconv}). 
We focus on the design and implementation of neuromorphic processing units based on our new \textit{NeuroAI TNN} (NeuTNN) microarchitecture, which aggressively exploit brain-like attributes for highly energy-efficient execution of future AI workloads.

This work presents a new NeuTNN design approach that builds on and extends previous work on TNNs by incorporating recent findings and insights from neuroscience. Specifically, this approach focuses on CMOS implementations of CCs composed of AD  neurons, adhering to the neocortical structure and functional hierarchy \cite{ahmad2016neurons,hole2021thousand}.
We incorporate fundamental tenets of TNNs and CCs, and build on microarchitecture model and optimizations from prior works to develop a new broader category of TNNs along with a toolsuite to enable the design of novel neuromorphic processors.  


To facilitate the design and implementation of NeuTNNs, we develop an automated PyTorch-to-layout tool suite called \textit{NeuTNNGen} consisting of a high-level functional simulator and an accompanying hardware generator. 
This work extends beyond TNNGen \cite{vellaisamy2024tnngen} to incorporate active dendrite neurons with more capable and efficient columns and layers, providing a much richer design space.
Our key contributions include:

\begin{itemize}
    \item \textbf{NeuTNN with Active Dendrite Neuron}: We introduce a new broader category of TNNs called NeuroAI TNN (NeuTNN) and demonstrate hardware implementation of NeuTNNs based on a novel neuron model with active dendrites featuring distal and proximal segments. This model significantly enhances computational efficiency and expressiveness compared to conventional point-neuron models in prior works.
    
    \item \textbf{NeuTNNGen Design Framework}:We develop an automated design framework based on the NeuTNN microarchitecture that translates PyTorch models to application-specific chip layout. Compared to prior TNNGen, NeuTNNGen has support for fully configurable multilayered networks and enhanced verification with automated compatibility checks for consistency of design parameters across hierarchical abstraction layers.

    \item \textbf{Implementation of Reference Frames}: To the best of our knowledge, this is the first work to show the feasibility of implementing RFs in CMOS with spiking neurons (NeuTNNs) and report their hardware results.




    \item \textbf{Post-Layout Power-Performance-Area (PPA)}:
    Post-layout PPA results for 45nm and predictive 7nm CMOS are reported for several benchmarks: time-series clustering (expanding on the number of benchmarks from \cite{vellaisamy2024tnngen}), MNIST, and Place Cells in RF. We leverage TNN7 \cite{nair2022tnn7} library to realize optimized NeuTNN implementations.
    
\end{itemize}

Rest of the paper is organized as follows. 
Section \ref{sec:neutnn} describes the NeuTNN microarchitecture model. Section \ref{sec:neutnngen} presents the NeuTNNGen design framework. Section \ref{sec:exp_framework} details the experimental setup used to analyze post-layout hardware complexity of the generated designs. Sections \ref{sec:results_UCR},  \ref{sec:results_MNIST},  \ref{sec:results_PC} present the experimental results for UCR benchmarks, MNIST, and Place Cells, respectively. Finally, Section \ref{sec:concl} summarizes key conclusions with suggestions for future work.

\section{NeuroAI Temporal Neural Networks (NeuTNN)}
\label{sec:neutnn}

\subsection{NeuTNN Microarchitecture and Organization}


\begin{figure*}[t]
\centering
\includegraphics[width=0.9\textwidth]
{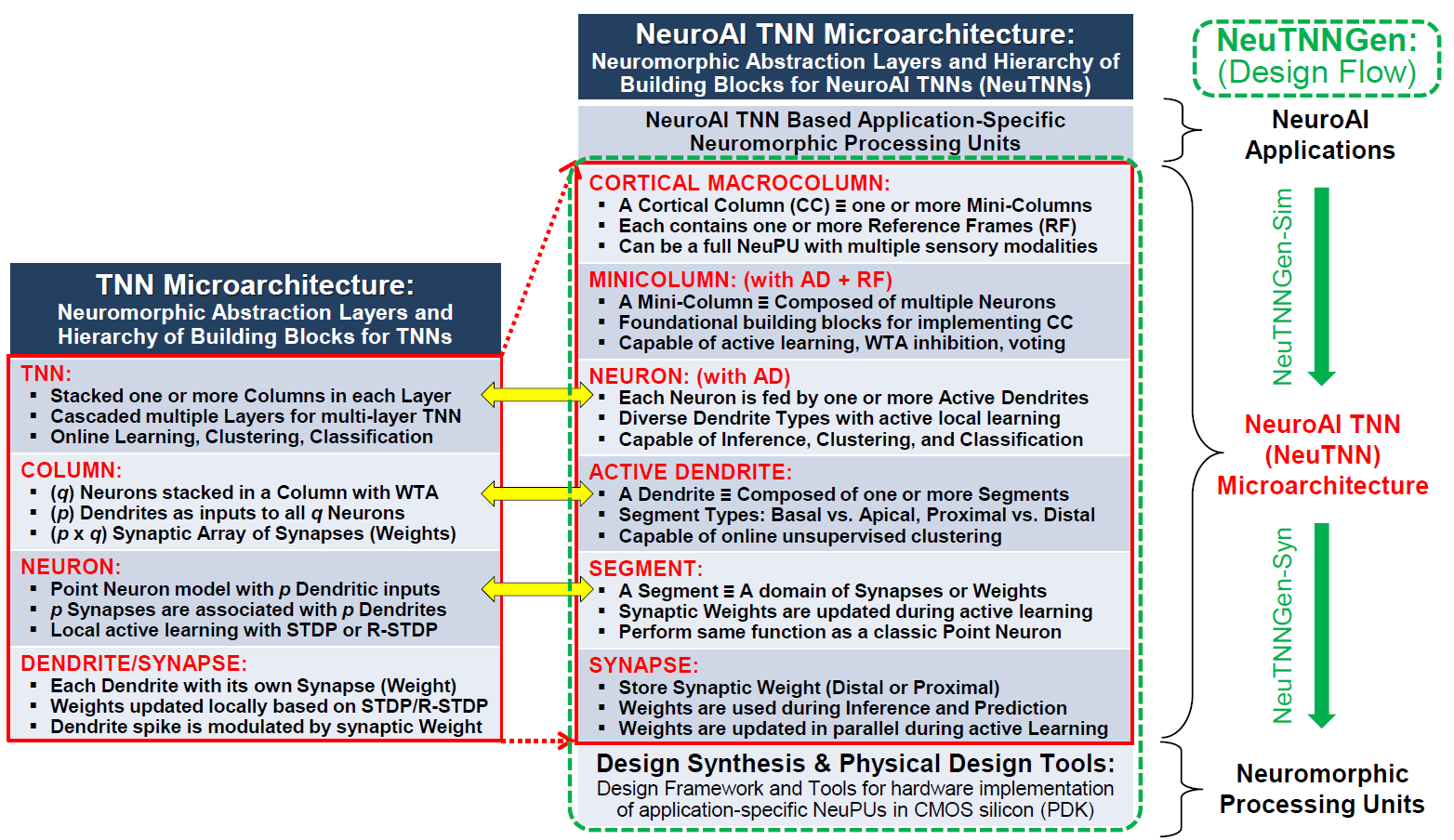}
    \caption{NeuTNNGen framework significantly extends beyond prior TNNGen \cite{vellaisamy2024tnngen} in functional modeling capabilities and design exploration and synthesis tools. The NeuTNN microarchitecture model consists of a hierarchy of six abstraction layers spanning from synapses up to Cortical (macro) Columns (CC), with support for Active Dendrites (AD).}
\label{fig:NeuTNN_microarchitecture}
\end{figure*}

\begin{figure*}
\centering
\includegraphics[width=0.8\textwidth]{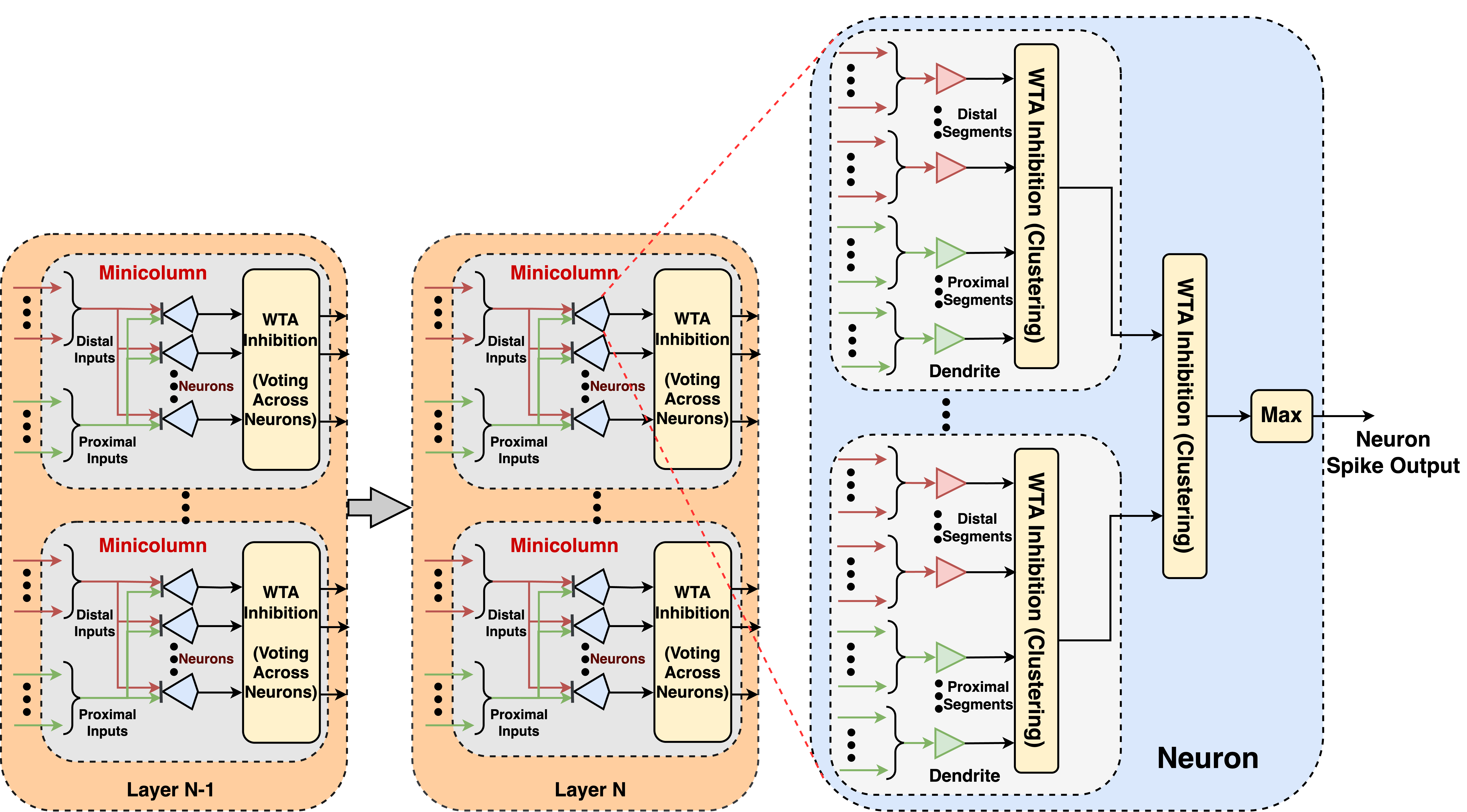} 
\caption{NeuTNN microarchitecture organization: Consist of one or more layers, wherein each layer consists of multiple minicolumns stacked in parallel. A minicolumn contains a group of neurons fed by a group of active dendrites. Each blue polygon is a neuron with one or more active dendrites, each containing one or more proximal and/or distal segments. Note that there are two levels of clustering just within a single neuron and three levels of clustering within a single minicolumn.}
\label{fig:neutnn_neuron}
\end{figure*}


NeuTNN microarchitecture (Fig. \ref{fig:NeuTNN_microarchitecture} and Fig. \ref{fig:neutnn_neuron}) is directly based on the biological hierarchy of abstraction layers and building blocks in the neocortex, incorporating new features from neuroscience. As shown in Fig. \ref{fig:NeuTNN_microarchitecture}, these abstractions are composed bottom-up as follows: 1) synapse that receives input and generates a weighted response, 2) segment that accumulates multiple synaptic responses, 3) active dendrite that performs input clustering via multiple distal and proximal segments, 4) neuron that performs contextual clustering and classification via multiple active dendrites, 5) minicolumn that performs higher-level functionalities such as voting and serves as the foundational building block for RFs, and 6) cortical macrocolumn with RFs that can be used to build NeuTNNs with diverse sensory modalities.

Fig. \ref{fig:NeuTNN_microarchitecture} also highlights the difference between the previous TNN model \cite{vellaisamy2024tnngen} and the new NeuTNN in terms of the neuromorphic abstraction layers and corresponding hierarchy of building blocks. The four layers of abstraction of TNN have expanded to six layers in NeuTNN.
Unlike TNN's classic point neuron, the NeuTNN neuron (Fig. \ref{fig:neutnn_neuron}) is significantly richer in functionality as it comprises active dendrites, each of which can contain multiple segments of different types (distal and proximal). Each NeuTNN segment is operationally equivalent to a classic TNN point neuron. 
A NeuTNN active dendrite is computationally comparable to a TNN column (see yellow bi-directional arrows in Fig. \ref{fig:NeuTNN_microarchitecture}).

As shown in Fig. \ref{fig:neutnn_neuron}, multiple NeuTNN neurons can be grouped to form a NeuTNN minicolumn, the foundational building block for NeuTNN. Minicolumns can be stacked into a large layer and multiple layers can be cascaded to form a multilayer NeuTNN. However, as a result of AD neuron model, a single NeuTNN layer can potentially replace a multilayer TNN, resulting in shallower networks. NeuTNNs  enable highly energy-efficient neuromorphic processing units.

\subsection{Functional Building Blocks of NeuTNN}

As mentioned earlier, NeuTNN involves a bottom-up hierarchy with building blocks in each abstraction layer used to form the next layer's building blocks.
NeuTNN incorporates distinct submodules for synaptic processing and neuronal integration. Synaptic units can employ a ramp-no-leak (RNL) or step-no-leak (SNL) response function for inference operations and incorporate STDP for synaptic learning (as in TNNs \cite{nair2021microarchitecture}). NeuTNN employs a much more robust neuron model that contains active dendrites with distal and proximal segments.
These segments receive and process synaptic input locally. Each segment operates similarly to a basic point neuron, but it utilizes different response characteristics.

The dendrite integrates the signals from its segments, selecting the contribution of the highest activated segment to influence the somatic potential. Multiple such dendrites compete to perform clustering and preprocessing of inputs before being fed to the neuron body. This neuronal design forms the basis for the hierarchical structure of NeuTNN. A minicolumn consists of a collection of interconnected neurons, each featuring multi-segment active dendrites. A full NeuTNN layer is composed of multiple minicolumns, allowing for parallel processing and functional specialization within the layer.

In its simplest form, a proximal segment can degenerate to a 1-bit enable signal (such as a supervised class label) for the neuron \cite{smith2023neuromorphic}.
A neuron with only one active dendrite and one proximal input acting as enable plays a special role and is referred to as a \textit{clustering voter (CV)} unit in \cite{smith2023neuromorphic}. 
These CV units can be organized into CV groups.
For supervised classification, the number of CV units within each CV group usually corresponds to the number of output classes or labels relevant to the classification task.


\section{NeuTNNGen Design Framework and Toolsuite}
\label{sec:neutnngen}

NeuTNNGen framework spans software-level modeling and hardware design synthesis. It supports rapid design space exploration of NeuTNN designs using PyTorch to find the optimal model for a specific application. The toolsuite then translates the software model to hardware implementation and leverages commercial EDA tools to obtain post-synthesis and post-place-and-route PPA for application-specific NeuTNN designs.

\subsection{NeuTNN Hardware Design Flow}

\begin{figure}[t]
\centering
\includegraphics[width=0.86\columnwidth]{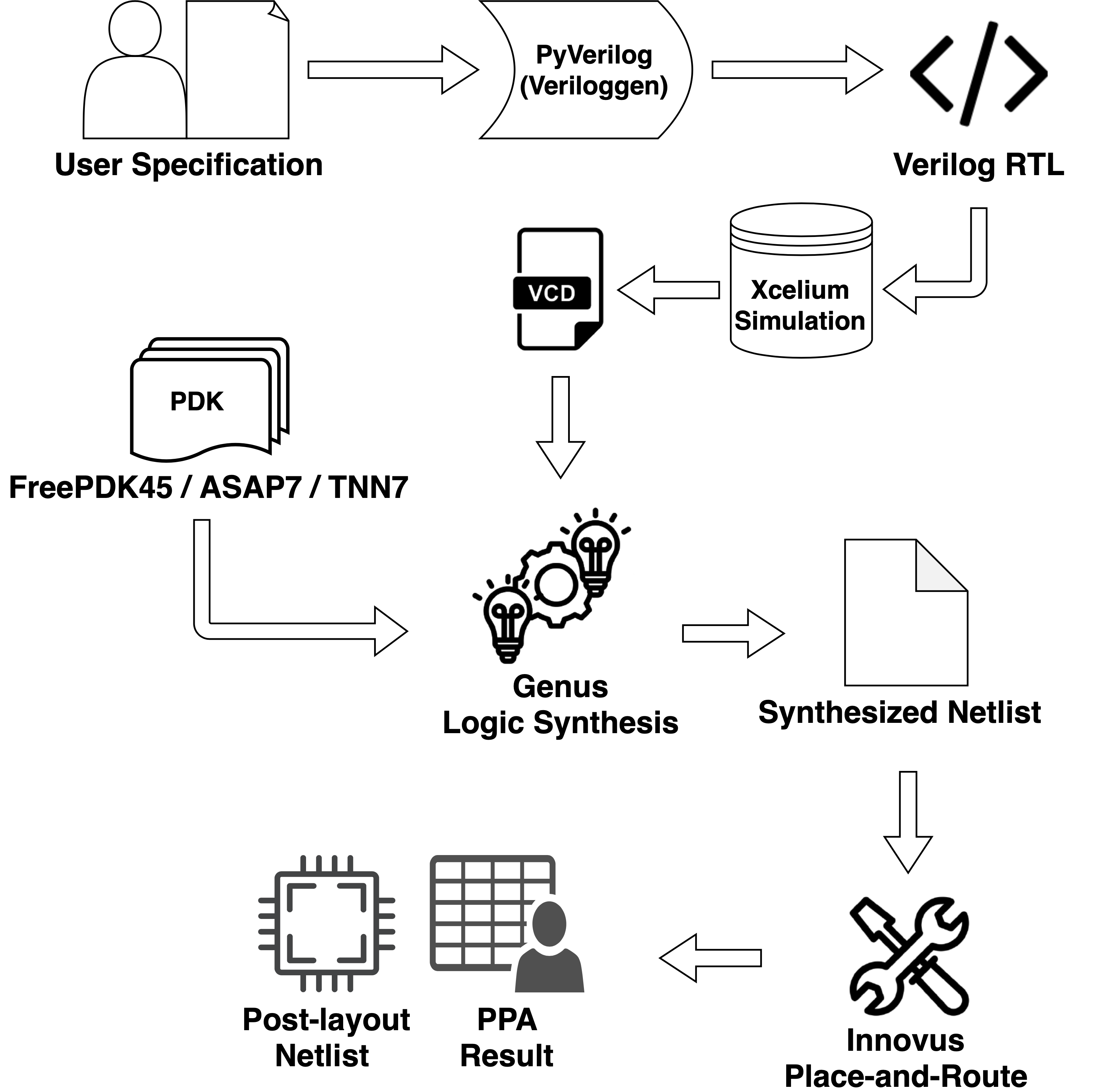} 
\caption{NeuTNNGen hardware process flow from RTL generation to layout:
Veriloggen converts software model to Verilog, which is simulated using testbench-driven vectors, producing switching activity file. 
This is sourced during Genus synthesis, along with standard cells and macros. 
Innovus uses the post-synthesis netlist to generate post-layout netlist and PPA.}
\label{fig:hardware_flow}
\end{figure}

Fig. \ref{fig:hardware_flow} portrays the NeuTNNGen hardware design workflow. User specifications (e.g., optimal layer parameters) are imported from PyTorch after functional performance simulation and translated to equivalent Verilog RTL via the Veriloggen package based on PyVerilog \cite{TakamaedaPyverilog}, providing users with an easily accessible Python interface similar to TNNGen \cite{vellaisamy2024tnngen}. All the NeuTNN abstractions from Fig. \ref{fig:NeuTNN_microarchitecture} are implemented as custom modules in PyVerilog. Subsequently, the generated RTL is subjected to functional verification via simulation using Cadence Xcelium. This step produces a Value Change Dump (VCD) file, capturing the switching activity essential for subsequent power analysis. Following verification, logic synthesis is performed using Cadence Genus. The synthesis process can target various standard cell libraries, including the custom TNN7 library \cite{nair2022tnn7} augmented with ASAP7, to generate a post-synthesis netlist. Finally, this netlist undergoes place-and-route using Cadence Innovus to produce the final post-layout netlist.



\subsection{User Configuration and Design Space Exploration}



User control over the NeuTNNGen workflow is primarily managed through a central configuration text file. This file requires users to define critical parameters before initiating any model generation process. Among the essential fields is \texttt{flow}, which dictates the specific hardware generation stages to be executed (e.g., RTL generation, RTL-level functional simulation, logic synthesis, and place-and-route). Users also specify the target technology node and associated PDK using the \texttt{node} argument, with options including NanGate45, ASAP7 or ASAP7 augmented with TNN7. Based on the selected node, the framework automatically generates appropriate technology-specific Tcl scripts from predefined templates for both synthesis (Cadence Genus) and place-and-route (Cadence Innovus).

When a layer is defined by the user, the framework automatically instantiates the necessary submodules based on the provided parameters (e.g., number of minicolumns, neurons, dendrites, distal/proximal segments, synapses, supervised/unsupervised STDP). Furthermore, each layer supports the integration of a kernel with configurable size and stride. Reference frame can be implemented using NeuTNN minicolumns as illustrated in \cite{smith2022macrocolumn}. The \texttt{Model} object sits at the top of the abstraction layers, serving as a wrapper for all user-defined layers. Through the \texttt{.add()} primitive, users can add layers sequentially. Upon invocation, NeuTNNGen checks if the input width of a new layer matches the output width of the last layer, ensuring valid connections between adjacent layers. Upon establishing compatibility with the rest of the model, it is instantiated with a unique identifier and appended to the model's internal list of layers. If a mismatch is detected at any interface, the framework provides feedback to the user, suggesting compatible parameter adjustments.


\subsection{Synaptic Pruning Optimizations}
\label{subsec:pruning}

Prior works \cite{nair2021microarchitecture, nair2022tnn7} have noted that synapses constitute majority of the TNN hardware complexity. As will be apparent from this work, same holds true for NeuTNNs. Hence, optimizing the synapse count is key to achieving area and power efficiency. This can be done via synaptic pruning. Another goal of such pruning is to introduce a form of regularization into NeuTNN, thereby improving model robustness and accuracy akin to dropout\cite{srivastava2014dropout} in traditional DNNs.

To facilitate such synaptic optimizations, NeuTNNGen incorporates support for synaptic pruning based on heuristics derived from weight profiling with minimal impact on model accuracy. These heuristics include histogram of weight distribution within each segment, dendrite, neuron, and layer. Users are then provided with the option to set unimportant synapses (i.e., weights with small magnitudes) to zero. Here, users can choose the range of weights that are deemed ``small" based on the weight histogram. After this step, users have the additional option to keep the remaining non-zero weights intact or set them to the maximum value, thereby making the weights bimodal or effectively 1-bit. After this pruning step is completed, the tool has two options: 1) disable incremental on-chip learning wherein the zero weights are not instantiated in hardware, significantly improving both area, leakage and dynamic power or 2) enable on-chip learning wherein the zero weights are still instantiated but initialized to zero and updated incrementally, improving dynamic power consumption. Due to NeuTNN's capability of online learning, any loss in model accuracy can be potentially recovered after model deployment on chip. 

\section{NeuTNNGen Experimental Framework}
\label{sec:exp_framework}

\subsection{Experimental Setup}
\label{sec:setup}

The NeuTNNGen framework supports multiple PDKs, including FreePDK45 \cite{oliveira2016ascend}, ASAP7 \cite{clark2016asap7}, and a combination of ASAP7 and TNN7 \cite{nair2022tnn7}, to generate post-layout netlists of diverse representative NeuTNN designs. As mentioned earlier, synthesis and place-and-route are done with Cadence Genus and Innovus, respectively, using automated Tcl scripts tailored for each stage of the design pipeline. Like TNNGen, NeuTNNGen does not support LVS verification post-place-and-route as it requires expert intervention. For consistency across the diverse designs for layout optimization, a constant floorplan density of 60\% is fixed. Adhering to biological timescale, 100 kHz clock frequency is used.

\subsection{Runtime Evaluations}
NeuTNN models with varying synapse counts and minicolumns are implemented to analyze the scaling trend across different designs and PDKs. Synthesis and place-and-route runtimes are collected to evaluate runtime speedups using ASAP7+TNN7 cell libraries. Process flow runs were performed on 2.4 GHz Intel(R) Xeon(R) E5-2640 v4 CPUs. 

\begin{figure}[t]
\centerline{\includegraphics[width=1.05\columnwidth]{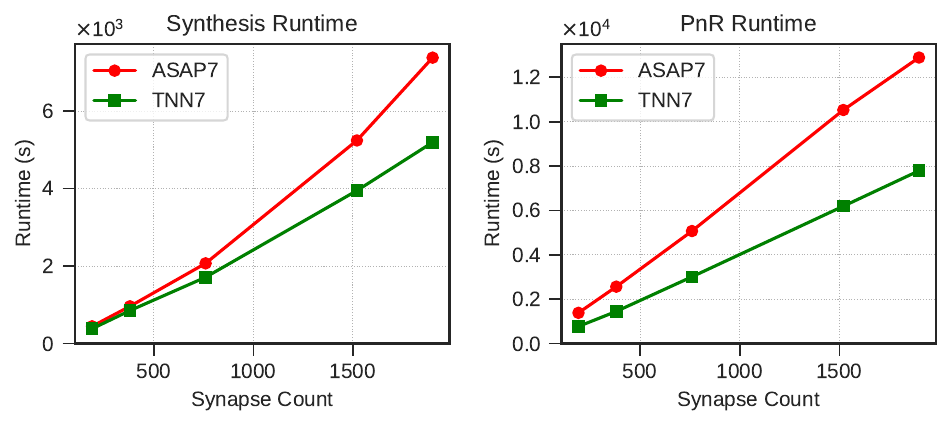}}
\caption{Synthesis (left) and  place-and-route (right) runtimes scaling  of NeuTNN design flows (for ASAP7 and TNN7 PDKs) with varying total synapse counts.}
\label{fig_runtime}
\end{figure}

Previous work on TNNGen \cite{vellaisamy2024tnngen} has shown the run-time scaling improvement of TNN7 over ASAP7. We confirm that NeuTNNGen maintains the same run-time scaling improvement for synthesis and place-and-route runtimes relative to increasing total synapse counts in the benchmark designs.
As seen in Fig. \ref{fig_runtime}, NeuTNNGen shows a consistent increase in runtime with increasing synapse count, with the runtime benefits of TNN7 (up to 40\%) over ASAP7 also increasing, aligning well with the observations from previous work on TNNGen \cite{vellaisamy2024tnngen}.



\begin{figure}[h!] 
    \centering 
    \begin{subfigure}[b]{0.46\textwidth} 
        \centering
        \includegraphics[width=0.97\textwidth]{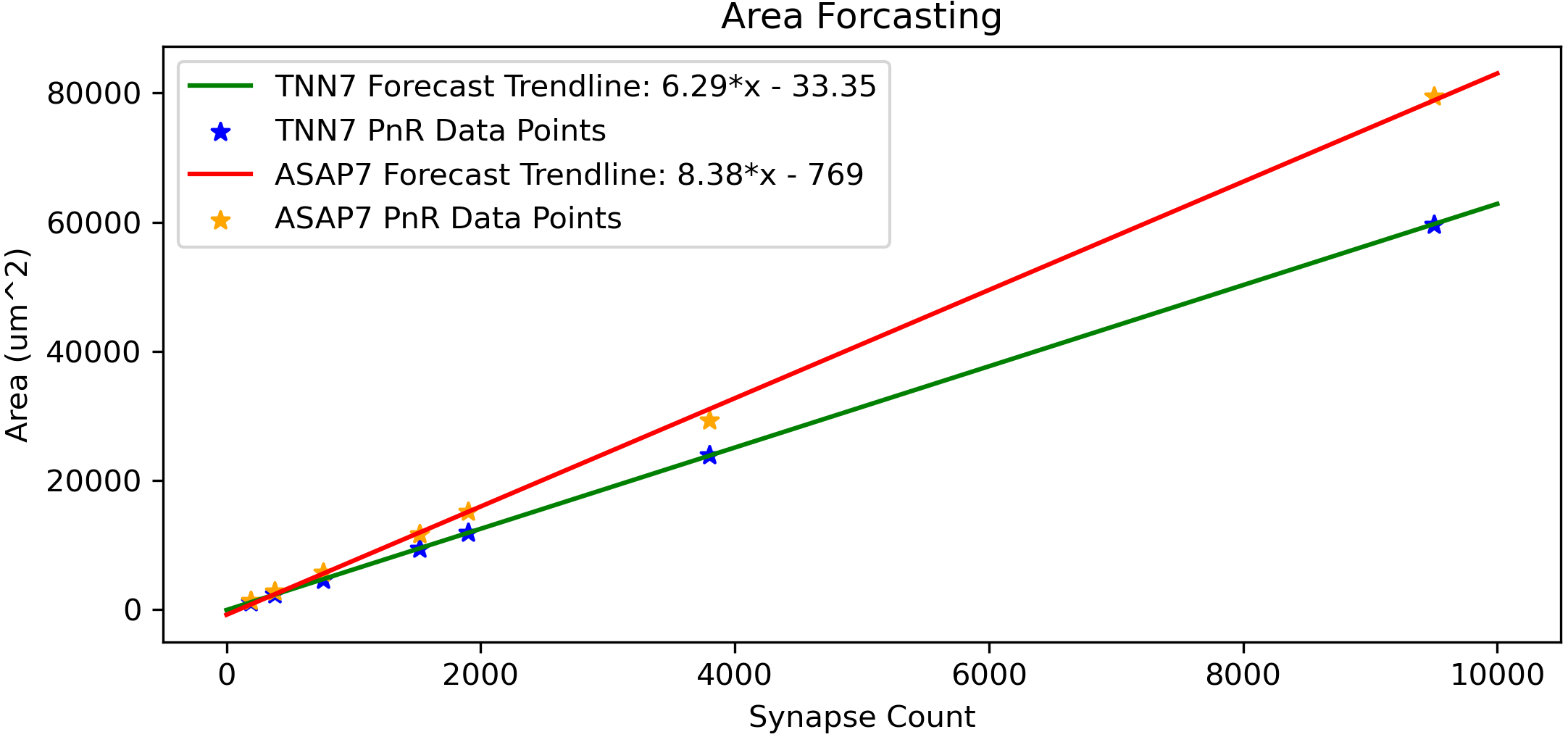} 
        \label{fig:area_forecast}
        \vspace{+4pt}
    \end{subfigure}
    
    \begin{subfigure}[b]{0.46\textwidth}
        \centering
        \includegraphics[width=0.94\textwidth]{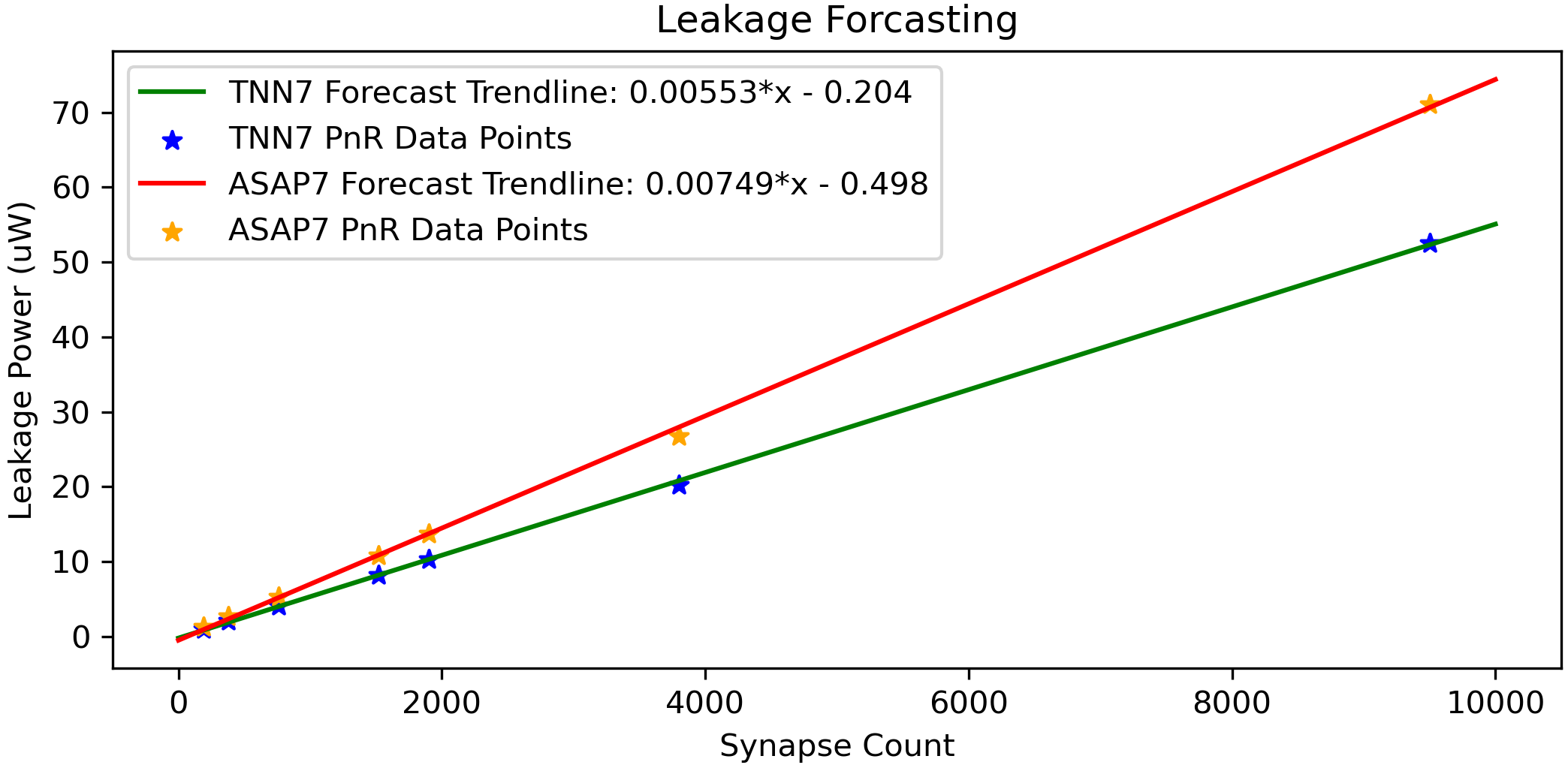} 
        \label{fig:leakage_forecast}
    \end{subfigure}
    \caption{Area and leakage power forecasting equations for NeuTNN designs for TNN7 and ASAP7  PDKs, based on results from actual benchmark designs. We see a clear linear scaling of area and power relative to synapse count for NeuTNN designs.} 
    \label{fig:fig_forecasts}
\end{figure}

\subsection{Area and Leakage Power Forecasting}


As evident from the previous section, synthesis and place-and-route can consume a significant amount of time especially for large designs. Even if runtime is not a concern, some users may not have access to industry EDA tools. Hence, we present a qualitative area and leakage power forecasting model for rapid design space exploration without the need to access or run any EDA tools, similar to TNNGen. The seven UCR benchmarks from TNNGen are used here to generate equivalent NeuTNN designs. Fig. \ref{fig:fig_forecasts} shows the area and leakage power trends for ASAP7 and TNN7 for NeuTNN designs. Note that TNN7 area and leakage power both scale desirably slower than ASAP7 with increasing advantage for larger designs. 
The corresponding forecasting equations are shown in the legend of Fig. \ref{fig:fig_forecasts}.

\section{Experimental Results on UCR Benchmarks}
\label{sec:results_UCR}

\subsection{Confirming Time-Series Benchmarks from TNNGen}

\begin{table}[t]
\centering
\caption{Post-place-and-route leakage power for NeuTNN designs for 12 representative UCR benchmarks (including the 7 from TNNGen \cite{vellaisamy2024tnngen}), using three different PDK cell libraries.}
\scalebox{0.96}{
\begin{tabular}{ccccc}
\hline
  \multicolumn{5}{c}{\textbf{Leakage Power for all three PDKs}} \\
\hline
 \textbf{UCR Benchmarks}& \textbf{Synapse}&  \textbf{FreePDK45} & \textbf{ASAP7} & \textbf{TNN7} \\
 \textbf{(SOTA Clustering)} & \textbf{Count} & \textbf{($m$W)} & \textbf{($\mu$W)} & \textbf{($\mu$W)} \\
\hline
SonyAIBORobotSurface2 & 130 & 0.32 &  0.94 &  0.74 \\
ECG200 & 192 & 0.45  & 1.39 & 1.06 \\
Wafer & 304 & 0.73 & 2.21 &  1.68 \\
TwoPattern & 512 & 1.22 & 3.79 &  2.85 \\
Coffee & 572 & 1.33 & 4.22 &  3.25 \\
ToeSegmentation2 & 686 & 1.66 & 4.92 & 3.79 \\
Plane & 1008 & 2.44 & 7.32 & 5.56 \\
Lightning2 & 1274 & 2.93 & 9.08 & 6.99 \\
Meat & 1344 & 3.02 & 9.57 & 7.42 \\
Beef & 2350 & 5.79 &  16.12  & 12.54 \\
OSULeaf & 2562 & 6.31 &  17.57  & 13.67 \\
WordSynonyms & 6750 & 19.8 & 49.37 & 40.27 \\
\hline
\end{tabular}
}
\label{tab:pnr_leakage_ucr} 
\end{table}


\begin{table}[t]
\centering
\caption{Post-place-and-route die area for NeuTNN designs for 12 representative UCR benchmarks (including the 7 from TNNGen \cite{vellaisamy2024tnngen}), using three different PDK cell libraries.}
\scalebox{0.93}{
\begin{tabular}{ccccc}
\hline
  \multicolumn{5}{c}{\textbf{Die Area for all three PDKs}} \\
\hline
 \textbf{UCR Benchmarks}& \textbf{Synapse}&  \textbf{FreePDK45} & \textbf{ASAP7} & \textbf{TNN7} \\
 \textbf{(SOTA Clustering)} & \textbf{Count} & \textbf{($\mu$m$^2$)} & \textbf{($\mu$m$^2$)} & \textbf{($\mu$m$^2$)} \\
\hline
SonyAIBORobotSurface2 & 130 & 15156.68 & 997.13 &  830.54 \\
ECG200 & 192 & 21884.31 & 1462.53 & 1200.73 \\
Wafer & 304 & 34769.06 & 2402.54 &  1919.28 \\
TwoPattern & 512 & 58558.41 & 4046.38 & 3232.47 \\
Coffee & 572 & 65420.72 & 4520.56 & 3611.28 \\
ToeSegmentation2 & 686 & 74064.40 & 5344.52 & 4426.00 \\
Plane & 1008 & 108829.33 & 7853.17 & 6503.51 \\
Lightning2 & 1274 & 138006.45 & 10419.41 & 8385.73 \\
Meat & 1344 & 145589.22 & 10991.91 & 8846.49 \\
Beef & 2350 & 262614.44 &  18178.67  & 15281.87 \\
OSULeaf & 2562 & 286305.61 &  19818.62  & 16660.49 \\
WordSynonyms & 6750 & 870555.73 & 54856.83 & 49478.29 \\
\hline
\end{tabular}
}
\label{tab:pnr_area_ucr} 
\end{table}
We first use NeuTNNGen to reproduce the hardware complexity results of TNNGen \cite{vellaisamy2024tnngen} targeting time series clustering benchmarks. In fact, we expand on the benchmark set used. Tables \ref{tab:pnr_leakage_ucr} and \ref{tab:pnr_area_ucr} present post-layout leakage power and die area results, respectively, for twelve representative time series benchmarks from UCR \cite{UCRArchive2018}. These benchmarks encompass diverse sensory modalities such as accelerometer, ECG, optical, RF sensors, etc. It can be seen that the NeuTNNGen results are very close to that of TNNGen, within 5\% error tolerance that could be caused due to cell placement and routing differences.
The largest design with a total of 6750 synapses incurs just 40.27 $\mu$W power and 0.049 mm\textsuperscript{2} area in TNN7. 
This shows that NeuTNNGen can still support the original TNN microarchitecture, performing highly efficient time series clustering within a few tens of $\mu$W power in 7nm CMOS.

\subsection{Synaptic Pruning Optimization for UCR Benchmarks}
We now present the ability of NeuTNNGen to enable design optimization by performing synaptic pruning, as presented in Section \ref{subsec:pruning}. Based on synaptic weight value profiling, we prune out synapses that are not contributory. Here, we identify synaptic weights with values less than half of the maximum weight value and set them to zero while setting the remaining weights to the maximum value, achieving 1-bit precision.

\begin{figure}[h!]
\centering
\includegraphics[width=\columnwidth, height=3.9cm]{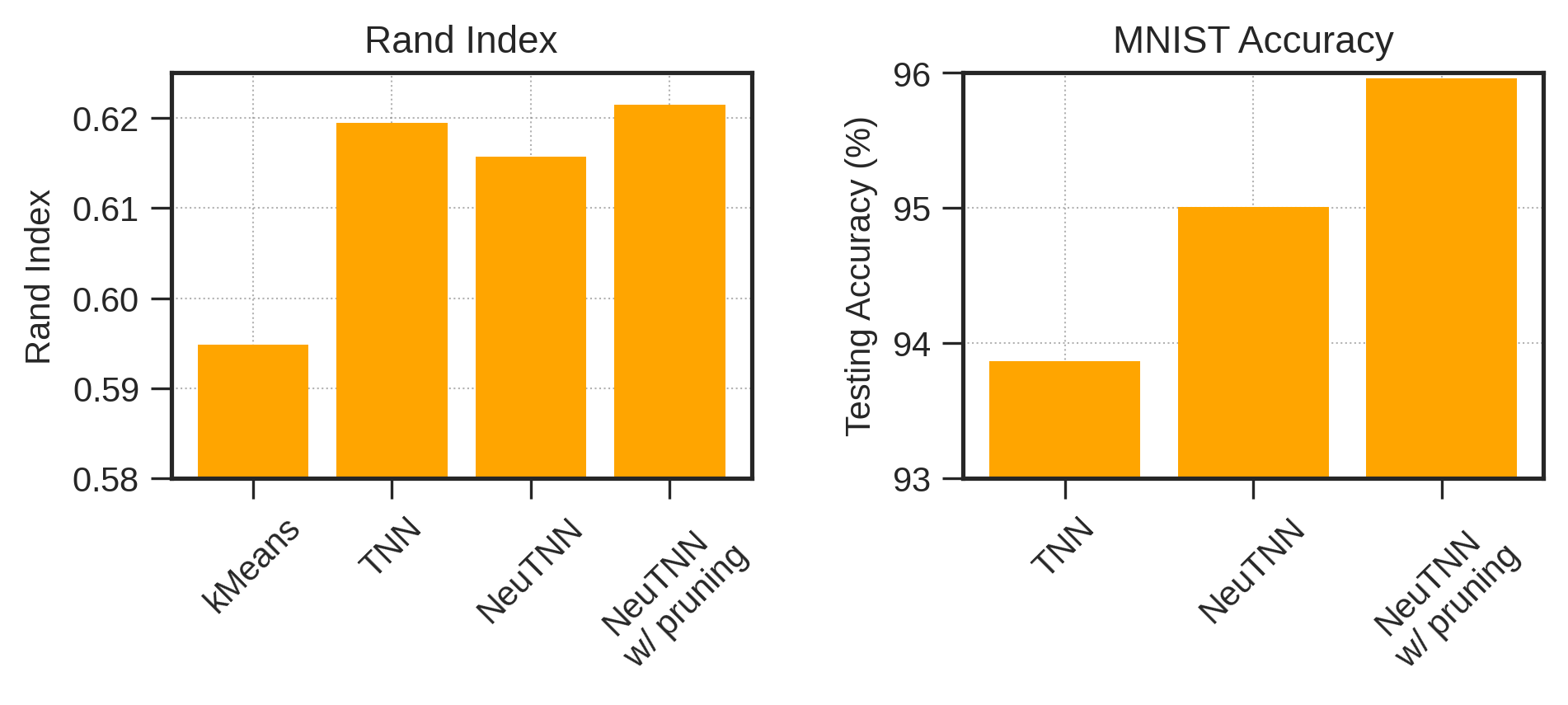}
    \caption{Comparison of design options across TNN, NeuTNN without pruning, and NeuTNN with synaptic pruning for UCR (left) and MNIST (right) benchmarks.}
\label{fig:mnist_acc_ucr_ri}
\end{figure}

Table \ref{tab:ucr_pruned_pnr_area_power} illustrates the potential optimization that can be achieved with synaptic pruning for the UCR benchmarks \cite{UCRArchive2018}. Across the 12 benchmarks, the synapse counts can be reduced by an average of 35\%, which results in similar reduction of power and die area by 34\% and 35\%, respectively. 
Fig. \ref{fig:mnist_acc_ucr_ri} illustrates that the Rand Index for pruned designs actually improved over the unpruned designs, suggesting possible negative interference involving some synaptic weights during training, which have been successfully pruned out. NeuTNN achieves slightly lower rand index compared to TNN as the design sizes curated for these benchmarks are too small to benefit the hierarchical representational power embodied by NeuTNN.


\begin{table}[t]
\centering
\caption{Post‑place‑and‑route power and die area results for NeuTNN designs from Tables \ref{tab:pnr_leakage_ucr} and \ref{tab:pnr_area_ucr} 
with synaptic pruning, for TNN7. The other two PDKs are omitted for brevity.}
\scalebox{0.95}{
\begin{tabular}{cccc}
\hline
  \multicolumn{4}{c}{\textbf{Optimized TNN7 Leakage Power and Die Area}} \\
\hline
 \textbf{UCR Benchmarks}& \textbf{Pruned}&  \textbf{Leakage Power} & \textbf{Die Area} \\
 \textbf{(SOTA Clustering)} & \textbf{Synapse Count} & \textbf{($\mu$W)} & \textbf{($\mu$m$^2$)} \\
\hline
\hline
SonyAIBORobotSurface2 & 113 & 0.64 & 722.57 \\
ECG200 & 146 & 0.81 & 912.55 \\
Wafer & 207 & 1.14 & 1305.11 \\
TwoPattern & 369 & 2.05 & 2327.38 \\
Coffee & 372 & 2.11 & 2347.33 \\
ToeSegmentation2 & 398 & 2.20 & 2567.08 \\
Plane & 625 & 3.45 & 4032.18 \\
Lightning2 & 828 & 4.54 & 5450.72 \\
Meat & 793 & 4.38 & 5219.43 \\
Beef & 1340 & 7.15 &  8710.67 \\
OSULeaf & 1409 & 7.52 &  9163.27 \\
WordSynonyms & 3983 & 23.76 & 29192.19 \\
\hline
\end{tabular}
}
\label{tab:ucr_pruned_pnr_area_power}
\end{table}


\section{Experimental Results on MNIST Application}
\label{sec:results_MNIST}

\subsection{NeuTNN Design for Online MNIST Application}

In this subsection, we extend beyond small single-layered TNNs and present post-layout power and die area of much larger NeuTNNs for the MNIST digit recognition benchmark. Using the MNIST benchmark as a vehicle, we demonstrate two key advances of NeuTNNGen over TNNGen: 1) support for multilayer network and 2) efficacy of the intrinsic hierarchy in active dendrite-based NeuTNN that enables it to achieve similar accuracy with reduced hardware complexity.

\begin{table}[t]
\centering
\caption{Post‑place‑and‑route results for three MNIST designs: 1) multi-layer TNN with no active dendrites, 2) single‑layer NeuTNN with active dendrites, and 3) single-layer NeuTNN with active dendrites and synaptic pruning, for TNN7. Results for the other two PDKs are omitted for brevity.}
\scalebox{0.98}{
\begin{tabular}{cccc}
\hline
  \multicolumn{4}{c}{\textbf{TNN7}} \\
\hline
 \textbf{MNIST NeuTNN Designs}& \textbf{Synapse}&  \textbf{Leakage Power} & \textbf{Die Area} \\
 & \textbf{Count} & \textbf{(mW)} & \textbf{(mm$^2$)} \\
\hline
\hline
Multi‑layer TNN      & 3,096,000 & 10.96 & 13.92\\
without AD & & & \\
\hline
Single‑layer NeuTNN  & 2,488,320 & 9.08 & 11.15\\
with AD & & &  \\
\hline
Single‑layer NeuTNN  & 1,343,692 & 5.01 & 6.15\\
w/ AD \& synaptic pruning & (pruned) & &  \\
\hline
\end{tabular}
}
\label{tab:mnist_pnr_area_power_}
\end{table}

Table \ref{tab:mnist_pnr_area_power_} presents the leakage power and die area comparisons of multilayer TNN (from \cite{smith2020temporal}) and single layer NeuTNN. 
The multilayer TNN with a total synapse count of 3,096,000, has 4 layers (ECCCVT from Table 3 in \cite{smith2020temporal}) achieving near 94\% accuracy on MNIST. 
The single layer NeuTNN for MNIST comprises 576 CV (clustering voter) groups, with 10 CV units (single-AD neurons) per group, totaling 2,488,320 synapses. This design with a lower synapse count achieves a better MNIST accuracy of 95\% (Fig. \ref{fig:mnist_acc_ucr_ri}).
Note that the goal here is not to present an MNIST-tuned design but to simply use MNIST to juxtapose the efficacies of large TNN and NeuTNN designs.

From Table \ref{tab:mnist_pnr_area_power_}, it can be seen that the NeuTNN MNIST design enables a reduction in synapse count of almost 20\%. This is due to the implicit hierarchical organization in NeuTNN that has significantly more representational power. Furthermore, the NeuTNN design also reduces both power and die area by about 18\% compared to the multilayer TNN MNIST design. 
This NeuTNN-based design with active dendrites achieves 95\% accuracy on MNIST (Fig. \ref{fig:mnist_acc_ucr_ri}) with just 9.08 mW leakage power and 11.15 mm\textsuperscript{2} die area, based on post-layout PPA results using TNN7. In addition, this MNIST NeuTNN is capable of performing on-chip online continuous learning.

To reduce synapse count further, we can apply similar synaptic pruning to the NeuTNN design, resulting in a total of only 1,343,692 synapses (46\% reduction). With synaptic pruning (Table \ref{tab:mnist_pnr_area_power_}), the power is reduced to 5.01 mW and die area to 6.15 mm\textsuperscript{2}, representing a corresponding improvement of 45\%. 
Interestingly, 
Fig. \ref{fig:mnist_acc_ucr_ri} shows that the pruned MNIST design actually achieves an accuracy of close to 96\%, better than the unpruned version. More research is needed on synaptic pruning, which is not the primary focus of this paper, but mainly used to illustrate the use of NeuTNNGen for the simulation and evaluation of promising design optimization ideas. 

\subsection{NeuTNN Design Tradeoffs for MNIST Application}


\begin{figure}[t]
\centering
\includegraphics[width=0.95\columnwidth]{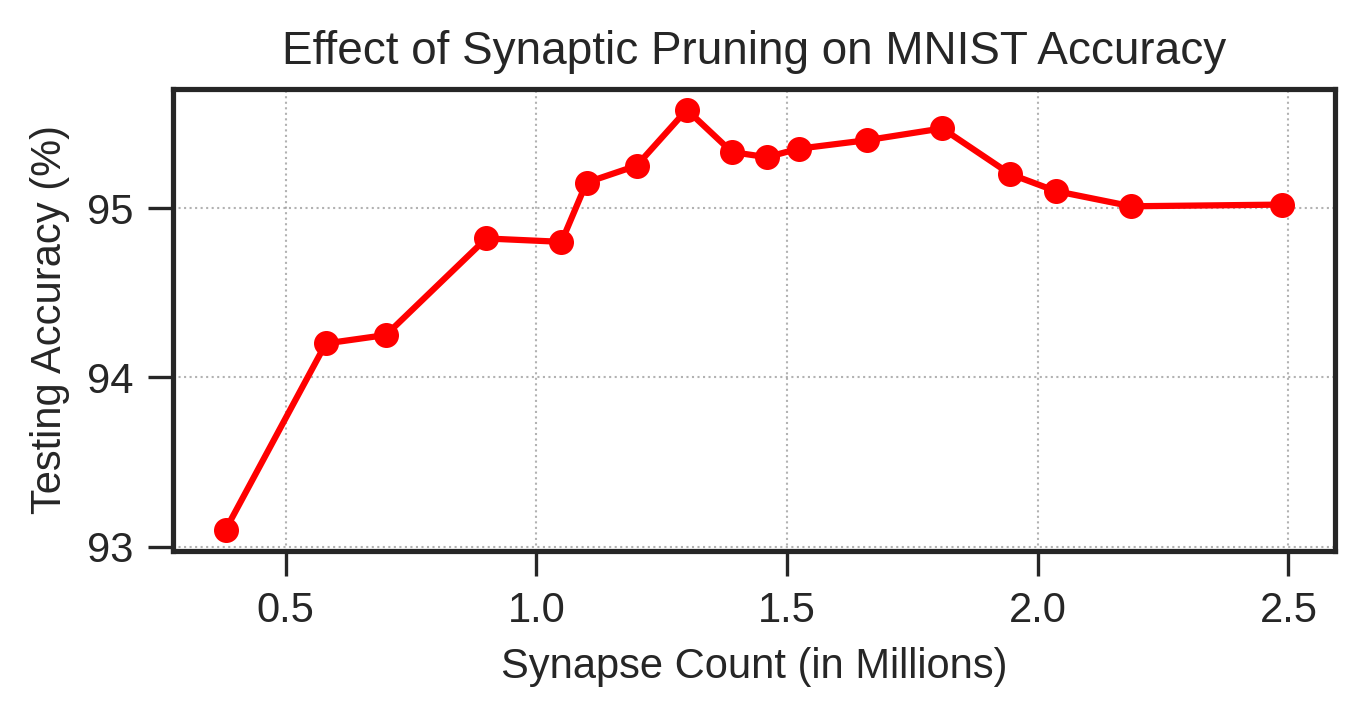}
    \caption{Design options using various synaptic pruning techniques mentioned in Sec.\ref{sec:neutnngen}.B with their MNIST accuracies. This demonstrates exploring design space using NeuTNNGen.}
\label{fig:synaptic_pruning_on_mnist_acc}
\end{figure}

Finally, we show that NeuTNNGen can be used to explore design space and trade-offs
for NeuTNN designs. Each NeuTNN abstraction layer has a set of functional building blocks that are parameterizable and composable. Building blocks from one layer are composed to form  next layer's building blocks, which are also parameterizable and composable. Design optimizations such as synaptic pruning can be applied across layers.
Hence, NeuTNNGen can rapidly explore various design choices. Fig.\ref{fig:synaptic_pruning_on_mnist_acc} shows possible MNIST NeuTNN designs as we prune synapses starting with the unpruned design with 2.5M synapses. As we prune, the MNIST accuracy initially improves gradually. Peak accuracy is achieved with synapse count of 1.3M, reflecting the optimal pruning point for the MNIST NeuTNN. Further pruning leads to loss of accuracy. However, even with only 0.25M synapses (10\% of the original 2.5M synapses), the accuracy drop is less than 3\%, indicating opportunity for significant pruning. NeuTNNGen is an effective tool for exploring and discovering optimal design points.

\section{Experimental Results on Place Cells Design}
\label{sec:results_PC}
 




\begin{figure}[t]
\centering
\includegraphics[width=1.0\columnwidth]{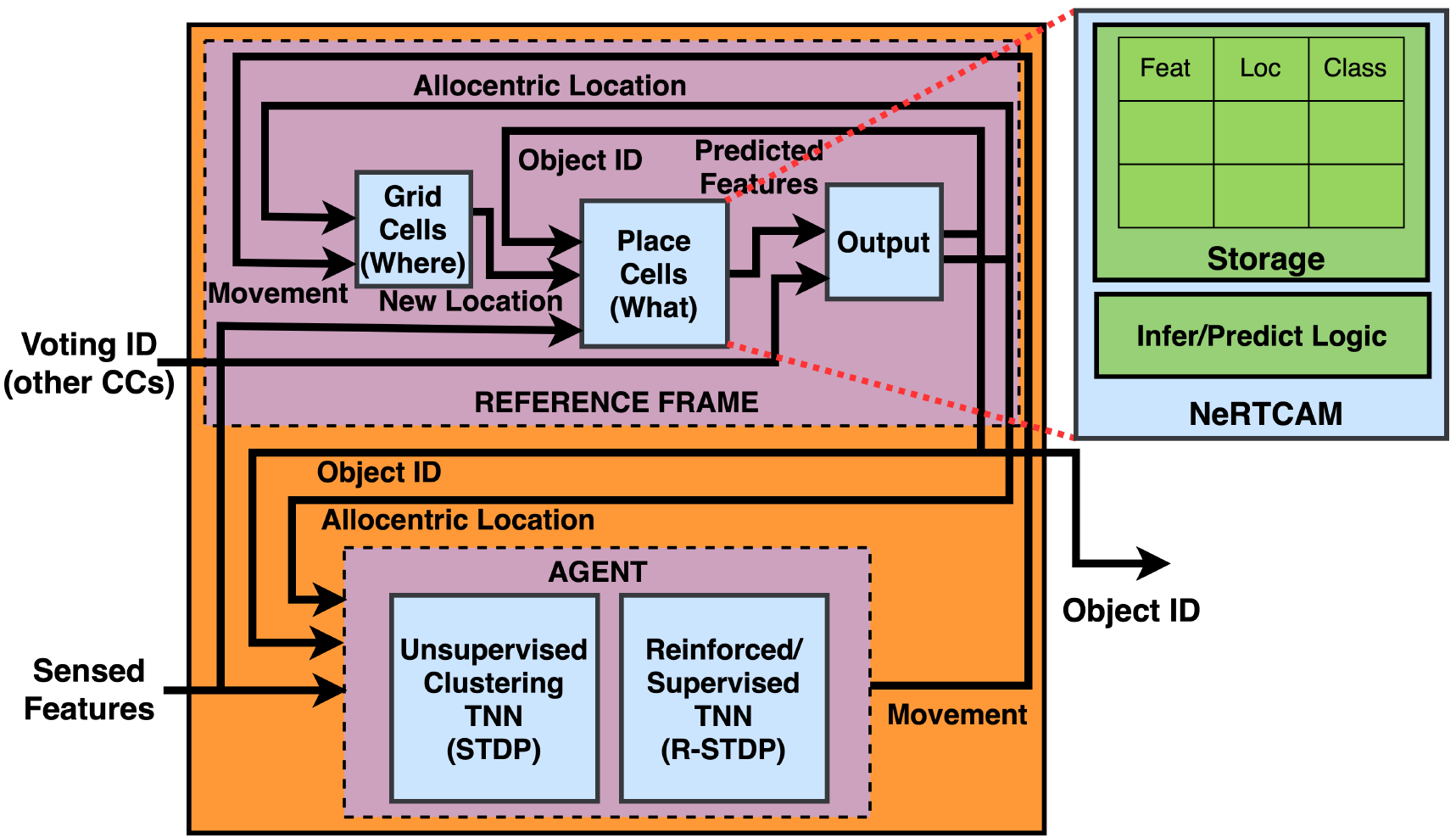}
    \caption{Cortical Column (CC) consists of a feedforward part called Agent and Reference Frame (RF) for memory \cite{shen2023cortical}. A key component of RF is \textit{Place Cells} which holds “sensory map”. NeRTCAM, proposed in \cite{nair2024nertcam}, is a CAM structure that implements the storage and inference/prediction logic in Place Cells. 
    Figure is taken from \cite{nair2024nertcam} with permission from authors.}
\label{fig:NeRTCAM_RF}
\end{figure}

In this section, we show how place cells (PC) in reference frames (RF) can be implemented as a specialized NeuTNN design using NeuTNNGen. Fig. \ref{fig:NeRTCAM_RF} illustrates the organization of a cortical column (CC) containing the feed-forward part called Agent and the RF to store memory from past learning. It also shows that PC can be implemented in a CAM-like structure called NeRTCAM \cite{nair2024nertcam}.

We use a design of place cells proposed in \cite{smith2022macrocolumn} as the third example of a NeuTNN design using NeuTNNGen. The place cells (PC), the key module within a reference frame (RF), can be built using three NeuTNN minicolumns, two of which share the same parameters while the third one differs. The unique parameters for the two minicolumn types are detailed in Table \ref{tab_placecell}. Two minicolumns with 388.8K synapses each and a third minicolumn with 454.4K synapses constitute the place cells of the RF with a total of 1,232K synapses. 

This design is demonstrated on a spatial navigation task with environments defined by a set of features at specific locations. The goal is to learn different environments as spatial feature-location maps stored in the place cell, orient in an unknown environment, and eventually navigate to a target location. The feature-location information is converted to spatial spike inputs appropriately routed to distal and proximal segments in the minicolumns.
Further details of the NeuTNN based design of place cells are omitted here for brevity (interested readers are referred to \cite{smith2022macrocolumn} for more details).



Table \ref{tab_placecell_complete} provides leakage power and die area for the full place cells NeuTNN design with three minicolumns. Each minicolumn employs multi-dendrite neurons. Combining the results presented in \cite{smith2022macrocolumn} with our PPA results in Table \ref{tab_placecell_complete}, highly efficient learning and effective orientation across 40 different environments (each 30x30 2D grid) can be achieved with just 7.26 mW leakage power and under 9.12 mm\textsuperscript{2} die area using TNN7. These results not only underscore the potential for stacking and cascading multiple minicolumns to form diverse NeuTNN organizations but also demonstrate the potential for designing highly practical and efficient NeuTNN designs for diverse edge-AI applications. NeuTNNGen can also be a great tool for exploring feasible implementations of reference frames, which is a wide open and promising research domain.

\begin{table}[t]
\caption{Place cell design from \cite{smith2022macrocolumn} is composed of three minicolumns: one large minicolumn (1x minicolumn \#1) and two small minicolumns (2x minicolumn \#2). Composition of each minicolumn in terms of number of neurons, dendrites, segments, and synapses are shown here.}
\centering
\scalebox{0.95}{
\begin{tabular}{cccccc}
\hline
\textbf{Minicolumn} & \textbf{Neuron} & \textbf{Dendrites} & \textbf{Segments} & \textbf{Synapses} & \textbf{Total} \\
\textbf{Type} & \textbf{Count} & \textbf{/ Neuron} & \textbf{/ Dendrite} & \textbf{/ Segment} & \textbf{Synapses} \\
\hline
\#1 & 40 & 10 & 16 & 71 & 454,400 \\
\#2 & 30 & 10 & 16 & 81 & 388,800 \\
\hline
\end{tabular}
}
\label{tab_placecell}
\end{table}

\begin{table}[t]
\caption{The entire Place Cells Implementation employs both types of minicolumns from Table \ref{tab_placecell}. Total Leakage Power and Die Area for the full Place Cells design, with a total of 1,232,000 synapses, are shown for all three PDKs.}
\centering
\scalebox{0.99}{
\begin{tabular}{crrr}
\hline
\textbf{Library} & \textbf{Synapse} & \textbf{Leakage Power} & \textbf{Die Area} 
\\
\textbf{(PDK)} & \textbf{Count} & \textbf{(mW)} & \textbf{(mm$^2$)} \\
\hline
FreePDK45 & 1,232,000 & 4157.32 & 185.76 \\
ASAP7 & 1,232,000 & 9.72 & 11.63 \\
TNN7 & 1,232,000 & 7.26 & 9.12 \\
\hline
\end{tabular}
}
\label{tab_placecell_complete}
\end{table}

\section{Summary and Future Work}
\label{sec:concl}

Based on this work, we can make some key observations.
(1) NeuTNNGen marks a significant step toward creating an automated framework that enables NeuroAI application developers to prototype neuromorphic processing units, incorporating recent neuroscience insights with active dendrites containing distal/proximal segments. 
(2) It is possible to create a toolchain for NeuTNN design that spans from PyTorch model to chip layout and allows exploration and design optimization such as synaptic pruning. 
(3) NeuTNNs can be highly performance efficient due to their rich hierarchical representational power. Our pruned single-layer NeuTNN for MNIST requires only 1.34M synapses (weights/parameters) with 96\% accuracy. A recent multilayer SNN design \cite{mozafari2019} for MNIST requires 36.7M synapses with 97\% accuracy. 
(4) The benefits of using TNN7 custom macro cells in NeuTNNGen continue to be quite significant in reducing the design time and design complexity.
(5) Implementing place cells of reference frame with NeuTNNGen yields promising initial results and demonstrates potential for future efficient implementation of NeuTNNs with complete cortical (macro)columns.


Several key areas need further development. 
(1) Potential applications of NeuTNNs involve diverse sensory modalities. Domains such as telecommunication, healthcare, automotive, factory automation, defense, and aerospace can benefit from low-power, edge-native, online-intelligent sensory processing. Diverse input encoders need to be added to support various sensor data types, including images, audio, videos, radio, acoustic, and other time series signals. 
(2) There is the potential to incorporate more brain-like architectures.
We are currently focusing on developing and incorporating complete reference frames \cite{hawkins2017theory} feature in NeuTNNGen. 
(3) Support for compute-in-memory \cite{nair2024tnn} and native CAM-based macros \cite{nair2024nertcam} can significantly improve hardware efficiency. 
(4) For more robust verification of new modules, assertion-based testing can be integrated into the set of current testing functionalities.

Going further, specialized neuromorphic or NeuroAI processing units can be created and integrated into existing mobile and datacenter Systems-on-Chip (SoCs) or emerging chiplet-based packages. 
For tighter integration, NeuTNNs can potentially be incorporated as specialized processing units within existing and emerging heterogeneous coupled architectures containing CPUs, GPUs,  NPUs, to enhance their performance and energy efficiency for diverse AI workloads. This work is our initial proof-of-concept effort. We expect the NeuTNNGen framework and tools to continue to evolve and mature.

\bibliographystyle{IEEEtran}
\bibliography{refs}

\end{document}